\documentclass[12pt]{article}
\usepackage{amsmath}
\usepackage{feynmp}
\def\e{{{\,\rm e}\,}}
\newcommand{\N}{\ensuremath{\mathcal{N}=4}~}
\newcommand{\Tr}{\ensuremath{\mathop{\mathrm{Tr}}}}

\newcommand{\sign}{\ensuremath{\mathop{\mathrm{sign}}}}
\newcommand{\AdS}{AdS$_5\times S^5$}
\newcommand{\path}{\ensuremath{\mathop{\cal P}}}

\begin{document}

\begin{titlepage} 
\begin{flushright}
ITEP-TH-13/00\\
hep-th/0003055\\\bigskip
March 2000
\end{flushright} 
\vspace{0.5cm} 
 
\begin{center} 

\textbf{\Large  Wilson Loops in
${\cal N}=4$ Supersymmetric Yang--Mills Theory}\\ 
\bigskip
{\large J.K.~Erickson$^1$, 
G.W.~Semenoff$\,^{3}$
and K.~Zarembo$^{1,2,4}$} \\
Email: \texttt{erickson@nbi.dk}, 
\texttt{semenoff@ias.edu}, 
\texttt{zarembo@physics.ubc.ca} \\ 
\bigskip
${}^1${\it Department of Physics and Astronomy}\\ 
${}^2${\it Pacific Institute for  Mathematical Sciences
\\University of British Columbia 
\\Vancouver, British Columbia, Canada V6T 1Z1}\\
\medskip
${}^3${\it Institute for Advanced Study, \\
Einstein Drive, Princeton, New Jersey 08540 USA\\}
\medskip
${}^4${\it Institute of Theoretical and Experimental Physics\\
B. Cheremushkinskaya 25, Moscow, 117218 Russian Federation}
\end{center} 
\bigskip
\begin{abstract} 
  Perturbative computations of the expectation value of the Wilson
  loop in ${\cal N}=4$ supersymmetric Yang-Mills theory are reported.
  For the two special cases of a circular loop and a pair of
  anti-parallel lines, it is shown that the sum of an infinite class
  of ladder-like planar diagrams, when extrapolated to strong
  coupling, produces an expectation value characteristic of the
  results of the AdS/CFT correspondence, $\langle
  W\rangle\sim\exp\bigl((\text{constant}) \sqrt{g^2N}\bigr)$.  For
  the case of the circular loop, the sum is obtained analytically for
  all values of the coupling. In this case, the constant factor in
  front of $\sqrt{g^2N}$ also agrees with the supergravity results. We
  speculate that the sum of diagrams without internal vertices is
  exact and support this conjecture by showing that the leading
  corrections to the ladder diagrams cancel identically in four
  dimensions.  We also show that, for arbitrary smooth loops, the
  ultraviolet divergences cancel to order $g^4N^2$.
\end{abstract}

\end{titlepage}

\newpage\setcounter{page}{1}
\begin{fmffile}{fmflong}

\section{Introduction and Summary}

The AdS/CFT correspondence conjectures a duality between \N
supersymmetric Yang-Mills theory in four dimensions and type IIA
superstring theory on an \AdS\ background
\cite{Mal97,Gub98,Wit98,Aha99,Akh99}. There are three levels to this
conjecture:
\begin{itemize}
\item In its strongest version the correspondence asserts that there
  is an exact equivalence between four dimensional \N supersymmetric
  Yang-Mills theory and type IIB superstring theory on the \AdS\ 
  background.  This also contains the conjecture that the \AdS\ 
  background is an exact solution of type IIB superstring theory.
\item A weaker version asserts a duality of the 't~Hooft limit of the
  gauge theory, where $N\rightarrow\infty$ with the 't~Hooft coupling
  $g^2N$ held fixed, and the classical $g_s\rightarrow 0$ limit of
  type IIA superstring theory on \AdS. In this correspondence,
  corrections to classical supergravity theory from stringy effects
  which are of order $\alpha'$ would agree with corrections to the
  large 't~Hooft coupling limit, of order $1/\sqrt{g^2N}$, but
  higher orders in $g_s$ on the supergravity side and non-planar
  diagrams on the gauge theory side could disagree.
\item An even weaker version is a duality between the 't~Hooft limit
  where one also takes the strong coupling limit $g^2 N
  \rightarrow\infty$ and the low energy, supergravity limit of type IIB
  superstring theory on \AdS. In this case, there would be order
  $\alpha'$ and $g_s$ corrections to supergravity which might not
  agree with order $1/N^2$ and $1/\sqrt{g^2N}$ corrections to \N
  supersymmetric Yang-Mills theory.
\end{itemize}
Even this last, weakest version of the correspondence has profound
consequences.  Previous to it, the only quantitative tool which could
be used to attack supersymmetric Yang-Mills theory was perturbation
theory in $g^2$, the Yang-Mills coupling constant.  This is limited
to the regime where $g^2$ is small.  Furthermore, although some
qualitative features of the large $N$ limit are known, it is not
possible to sum planar diagrams explicitly.  The AdS/CFT
correspondence enables one to compute correlation functions in the
large $N$, large $g^2 N$ limit.  This limit contains the highly
nontrivial sum of all planar Feynman diagrams, and emphasizes those
diagrams which have infinitely many vertices.

Because they are inaccessible to perturbation theory, the predictions
of the AdS/CFT correspondence are very difficult to check in any
direct way.  The main evidence which supports the correspondence comes
from symmetry arguments.  The global symmetries of both \N
supersymmetric Yang-Mills theory and type IIB string theory on \AdS\ 
are identical.  They have the same global super-conformal group
$SU(2,2\mid 4)$ (whose bosonic subgroup is $SO(4,2) \times SU(4)$).
Not only are the global symmetries the same, but some of those objects
which carry the representations of the symmetry group---the spectrum
of chiral operators in the field theory and the fields in supergravity
theory---can, to some degree, be matched \cite{Wit98}.  Furthermore, both
theories are conjectured to have an Montonen-Olive $SL(2,Z)$ duality
acting on their coupling constants.

The only correlation functions which can be checked directly are 
correlators which are protected by non-renormalization theorems and
thus do not depend on the coupling constant.  These correlators are
related to anomalies in the $SU(4)$ R-symmetry and conformal current
algebras.  A number of them have been shown to match \cite{Aha99}.

In this paper, we sum infinite classes of planar ladder diagrams in \N
supersymmetric Yang-Mills theory, and compare our results with the
predictions of the AdS/CFT correspondence. We are for the most part
interested in the Wilson loop operator. All computations are in the
't~Hooft limit where $N$ is put to infinity while holding
$g^2N$ fixed. We shall not succeed in providing a direct test of the
correspondence.  However, we will show that the generic dependence of
the loop expectation value on the coupling constant found in the
AdS/CFT computations arises naturally when perturbation theory is
summed to all orders and extrapolated to strong coupling.  We
also compute the exact weak coupling results to order $g^4N^2$ which,
if one takes the strongest version of the AdS/CFT correspondence
seriously, should tell us something about classical superstrings on
\AdS\ in the strong curvature limit.  We shall also find that leading
order corrections to the sum of ladder diagrams vanish for the two
simple loops that we study.

\subsection{The Wilson Loop}

The Wilson loop operator is a phase factor associated with the
trajectory of a heavy quark in the fundamental representation of the
gauge group, which in our case is $U(N)$.  The loop operator which
couples to classical quantities in the superstring theory in the
simplest way provides a source for a classical string
\cite{Mal98,Rey98,Dru99,Son00}. It is
\begin{equation}
\label{wilson}
W(C)=\frac{1}{N}\,\Tr\path\exp\oint_C d\tau\,\bigl(iA_\mu(x)\dot{x}_\mu
+\Phi_i(x)y_i\bigr),
\end{equation} 
where $x_\mu(\tau)$ is a parameterization of the loop, 
$y_i=\sqrt{\dot{x}^2}\theta_i$ and $\theta_i$ is a point on the
five dimensional unit sphere (${\theta^2=1}$).
This operator measures the holonomy of a heavy W-boson whose mass results from
spontaneous breaking of $U(N+1)$ gauge symmetry to $U(N)\times U(1)$.  In the
AdS/CFT correspondence, it is computed by finding the area of the world-sheet
of the classical string in \AdS\ whose boundary is the loop $C$, which in turn
lies on the boundary of AdS$_5$ \cite{Mal98,Rey98,Aha99}.

The expectation value of the Wilson loop operator is easy to compute
in perturbation theory
\begin{equation}
\langle W(C)\rangle=1+\frac{g^2N}{4\pi^2}\oint_C d\tau_1\,d\tau_2\frac{\lvert\dot
x(\tau_1)\rvert \lvert \dot x(\tau_2)\rvert-{\dot x}(\tau_1)\cdot \dot x(\tau_2) }{ \lvert
x(\tau_1)-x(\tau_2)\rvert^2} +\cdots.
\label{leadloop}
\end{equation}
For a loop without cusps or self-intersections, this result is finite.
This is because a cancellation occurs between the contributions of the
scalar and vector fields.  The case of cusps and self-intersections
has been discussed in \cite{Dru99}.

\subsection{Summary of results}

We work in Feynman gauge and consider specific, not necessarily gauge
invariant, classes of planar Feynman diagrams.  We will comment on the
issue of gauge invariance later.  In all computations, we consider
only planar diagrams, thus the coupling constant dependence of all
results is in the combination $g^2N$.

\subsubsection{Cancellation of ultraviolet divergences}

\N supersymmetric Yang-Mills theory is a conformal field theory for
any value of its coupling constant.  This conformal invariance is a
result of the fact that the theory has no dimensional coupling
constants, so it is conformally invariant at the tree level.  In
addition, the high degree of supersymmetry leads to a cancellation of
loop corrections to the coupling constant renormalization.

However, as we shall see, the theory does have divergent wave-function
renormalization, which does not contradict ultraviolet finiteness,
since all divergences can be absorbed by rescaling the
fields\footnote{In fact, some divergences are necessary in order for
  the theory to solve the conformal bootstrap equations
  \cite{Che93}.}.  The Wilson loop potentially contains its own
short-distance singularities which occur when the spacetime arguments
of operators in the loop approach each other.  In (\ref{leadloop}),
the divergent contributions of the gauge fields and the scalars
mutually cancel.  We shall see in the following that, to order
$g^4N^2$, part of such singularities survive in order to compensate
for the infinite wave function renormalization, and all ultraviolet
singularities cancel for a generic smooth (without cusps and
self-intersections) loop.

\subsubsection{Parallel lines}

The Wilson loop operator for a single infinite straight line, $W(C)$
where $C$ is parameterized by $x(s) =(s,0,0,0)$, is a BPS object: it
commutes with half of the sixteen supercharges \cite{Dru99}.  This
residual supersymmetry protects it from obtaining quantum corrections.
It is easy to verify that the first few orders in weak coupling
perturbation theory cancel.  In the strong coupling limit of planar
diagrams, the AdS/CFT correspondence predicts that
%\footnote{This is obtained by
%  taking the infinite $T$ and $L$ limit of the rectangular loop which
%  is studied in detail in section 5.}
\begin{equation}
\langle W(C)\rangle=1.
\label{infstrline}
\end{equation}
In fact, this is the case for any array of parallel straight lines with like
orientations.  

\subsubsection{Circular loop}

For the circular loop, conformal invariance predicts that the
expectation value of the loop operator is independent of the radius of
the loop\footnote{Note the subtle fact that, since the circle could
be obtained from an infinite straight line by a conformal
transformation, and this conformal transformation should be a symmetry
of the vacuum, the circular loop should have expectation value equal
to that of the infinite straight line (\ref{infstrline}).  We shall
see explicitly that this is not the case.  This asymmetry is
introduced by the boundary conditions which require that propagators
go to zero at infinity.  }.  We find that the sum of all planar Feynman
diagrams which have no internal vertices (which includes both rainbow
and ladder diagrams) produces the expression
\begin{equation}
\left< W(C)\right>_{\rm ladders}=
\frac{2}{\sqrt{g^2N}}\,I_1(\sqrt{g^2N}),
\label{circladder1}
\end{equation}
where $I_1$ is the Bessel function.  Taking the large $g^2N$ limit
gives 
\begin{equation}
\langle W(C)\rangle_{\text{ladders}}
=
\frac{\e^{\sqrt{g^2N}}}{(\pi/2)^{1/2}(g^2N)^{3/4}},
\label{circpert1}
\end{equation}
which has exponential behavior identical to the prediction of the
AdS/CFT correspondence \cite{Dru99,Ber98},
\begin{equation}
\left< W(C)\right>_{\rm AdS/CFT}
=
\e^{\sqrt{g^2N}}.
\label{circpert2}
\end{equation}

It is intriguing that this sum of a special class of diagrams produces
the exact asymptotic behavior that is predicted by the AdS/CFT
correspondence, considering that it does not include any diagrams
which have internal vertices.  Assuming that the AdS/CFT prediction is
indeed the correct asymptotic behavior, there are two possible reasons
for this agreement. First, the corrections to the sum of planar ladder
diagrams could produce a term which would be added to
(\ref{circladder1}) and which would grow no faster than
$\exp(\sqrt{g^2N})$ for large $g^2N$.  Such a term could modify the
prefactor of the exponential but would not modify the exponent.
Second, it is possible that corrections to the sum of ladder diagrams
cancel and the result (\ref{circladder1}) is exact.

In order to explore this appealing possibility, we compute the leading
order corrections to (\ref{circladder1}) coming from diagrams with
internal vertices.  These occur at order $g^4N^2$.  We find that these
diagrams do indeed cancel exactly when the spacetime dimension is
four.  Away from dimension four, there is a residual term of order
$(D-4)g^4N^2$.  It is tempting to speculate that all higher order
corrections from diagrams with internal vertices also cancel.  At this
point we have not been able to check this possibility beyond order
$g^4N^2$.  One reason we have to be optimistic is that the circular
loop is related to the single straight line by a conformal
transformation.  The conformal transformation must not be
representable as a unitary operator operating on the loop operator.
Otherwise, the expectation value of the circular loop would be one
(identical to the straight line) and we know this is not the case.
However, for some classes of diagrams, the symmetry could survive.  We
know that, for the straight line, the rainbow diagrams, which are what
the ladder diagrams are mapped onto, cancel identically.  This clearly
does not happen for their conformal images on the circle.  However it
could still happen for the diagrams which have internal
vertices---which do in fact cancel identically for the straight line
Wilson loop.  As stated above, we have checked explicitly that this is
indeed the case for the leading order $g^4N^2$ corrections.  This has
not yet been checked at higher orders.

One important issue is gauge invariance.  The computation of
(\ref{circladder1}) is in Feynman gauge. Since only a subset of the
diagrams has been included, there is no guarantee that the
calculation is gauge invariant. In fact, summing diagrams without
internal vertices in a different gauge would give a different result.
However, if the corrections indeed vanish then the sum would be equal
to the gauge invariant sum of all Feynman diagrams on the Wilson loop.
The cancellation of the diagrams with internal vertices would then be
a special property of the Feynman gauge\footnote{In Feynman gauge, the
  vector propagator coincides with the scalar one, which makes, for
  instance, cancellation of ultraviolet divergences more transparent
  than in other gauges: the second term in (\ref{leadloop}) vanishes
  even before integration along the loop.}.  We have explicitly
demonstrated that corrections vanish up to order $g^4N^2$.

\subsubsection{Anti-parallel lines}

Consider a rectangular Wilson loop with length $T$ and width $L$.  In
the limit that $T\gg L$, the ends can be ignored and the loop can be
seen as a pair of anti-parallel lines separated by a distance $L$.  We
find that the sum of all planar diagrams without internal vertices,
the ladder diagrams, is given by
\begin{equation}
\langle W(C)\rangle_{\rm ladders}=
\exp\biggl[\biggl(\frac{g^2N}{4\pi}-\frac{g^4N^2}{8\pi^3}\ln \frac{1}{g^2N}
+\cdots\biggr)\frac{T}{L}\biggr],\qquad g^2N\ll 1; 
\label{asympt1}
\end{equation}
and by
\begin{equation}
\langle W(C)\rangle_{\rm ladders}=
\exp\biggl[\biggl(\frac{\sqrt{g^2N}}{\pi}-1+{\cal O}\Bigl(\tfrac{1}{\sqrt{g^2N}}
\Bigr)
\biggr)\frac{T}{L}\biggr],\qquad g^2N\gg 1.
\label{asympt2}
\end{equation}
Here, we use the fact that $T\gg L$ to ignore any terms in the exponent which
are of lower order than $T/L$.  

The logarithm in the exponent of the weak coupling limit
(\ref{asympt1}) comes from an infrared divergence which appears in the
order $g^4N^2$ ladder diagram.  It was shown in \cite{Eri99} that
resummation of these logarithms to all orders replaces the logarithm
of $T/L$ by a logarithm of the coupling constant.  We will elaborate
on those arguments in section 5.  Logarithms of this kind are known to
appear in the Wilson loop in QCD at order $g^6$ \cite{appel}.  In the
weak coupling limit, the leading term and the coefficient of the
next-to-leading term are indeed gauge invariant.  However, the strong
coupling limit depends on the gauge parameter and we have quoted the
result (\ref{asympt2}) in Feynman gauge only (while the particular
constants vary, the generic behavior is still $\e^{\sqrt{g^2N}}$).

In the large 't~Hooft coupling limit, the AdS/CFT
computation\footnote{Note that the $g^2$ in \cite{Mal98} differs by a
  factor of $2$ from our definition.} \cite{Mal98,Rey98}
\begin{equation}
\langle W(C)\rangle_{\text{AdS/CFT}}=
\exp\Bigl(\frac{4\pi^2\sqrt{g^2N}}{\Gamma^4(1/4)}
\frac{T}{L}\Bigr),\qquad g^2N\gg1.
\label{adsasymp2}
\end{equation}
Comparing this with (\ref{asympt2}), we see that summing only ladder
diagrams does not produce (\ref{adsasymp2}). The failure of this
result to agree with the supergravity computation is not very
interesting: since the anti-parallel lines are not as closely related
to a BPS object as a circle, we have no reason to speculate that the
sum of ladder diagrams would reproduce the exact strong-coupling
behavior.

We also find that the diagrams with internal vertices
cancel exactly to the leading order $g^4N^2$.  In this case, as
opposed to the circular loop where they only cancel in four
dimensions, the cancellation occurs in any number of spacetime
dimensions.

The sum over an infinite number of diagrams that we have done produces
the strong coupling behavior
$\exp\bigl[(\text{constant})\sqrt{g^2N}\bigr]$ which is characteristic
of the Wilson loops computed using the AdS/CFT correspondence.  This
seems to be the generic behavior of infinite sums of planar ladder
diagrams.

\subsection{Outline}

In the rest of this paper we give a detailed derivation of the above
results.  In section 2, we give the results of some loop calculations.
In section 3, we prove that ultraviolet singularities cancel to order
$g^4N^2$ for any smooth loop.  In section 4, we consider the circular
loop.  We compute the sum over ladder diagrams and show that the
diagrams with internal vertices cancel to order $g^4N^2$. In section
5, we give similar arguments for two anti-parallel lines. In section
6, we summarize some speculations about the implications of our
results. In appendix A, we summarize our notation and conventions and
record some useful formulae.

\section{Perturbation theory}

With the exception of section \ref{canl}, we use regularization by
dimensional reduction throughout this paper.  This procedure considers
supersymmetric Yang-Mills theory in $2\omega$ dimensions as a
dimensional reduction of ${\cal N}=1$ supersymmetric Yang-Mills theory
in ten dimensions.  In this scheme, the gauge field $A_\mu^a(x)$ is a
$2\omega$ component vector field.  The index of the scalar field runs
over $10-2\omega$ values, $i=1,...,10-2\omega$.  In every dimension,
the fermion field has sixteen real components. Regularization by
dimensional reduction preserves the sixteen supersymmetries of the ten
dimensional Yang-Mills theory.  These lead to four conserved four
component Majorana spinor supercharges in four dimensions.  This
regularization scheme provides a supersymmetric regularization of the
four dimensional theory.  Since the gauge coupling becomes dimensional
in any spacetime dimension other than four, the regularization breaks
conformal symmetry explicitly.  It also modifies the R-symmetry.

\subsection{One loop self energy of the vector and scalar fields}

Consider the one loop self-energies of the vector and scalar field.
The dimension of space-time is $D=2\omega$.  All of the loop integrals
are elementary and can be found in reference books such as
\cite{Ram89}. We parameterize the Wilson loop by $x(\tau)$, and
abbreviate $x^{(i)}=x(\tau^i)$.

The vector field obtains self-energy corrections from:
\begin{itemize}
\item{}$N^2$ colors of vector fields and ghost fields:
\begin{multline*}
\parbox{20mm}{
\begin{fmfgraph}(20,15)
\fmfleft{i}
\fmfright{o}
\fmf{wiggly}{i,v1}
\fmf{wiggly,left,tension=.3}{v1,v2}
\fmf{wiggly,right,tension=.3}{v1,v2}
\fmf{wiggly}{v2,o}
\end{fmfgraph}}
\,\,+\,\,
\parbox{20mm}{
\begin{fmfgraph}(20,15)
\fmfleft{i}
\fmfright{o}
\fmf{wiggly}{i,v1}
\fmf{dots,left,tension=.3}{v1,v2}
\fmf{dots,left,tension=.3}{v2,v1}
\fmf{wiggly}{v2,o}
\end{fmfgraph}}\,
\\=
\delta^{ab}g^4N\frac{\Gamma(2-\omega)
\Gamma(\omega)\Gamma(\omega-1)}{(4\pi)^{\omega} \Gamma(2\omega)}\cdot
2(3\omega-1)\frac{ \delta_{\mu\nu}-p_\mu p_\nu/p^2}{p^{2-2\omega}}
\end{multline*}
\item{}$10-2\omega$ real scalar fields in the adjoint representation:
\[
\parbox{20mm}{
\begin{fmfgraph}(20,15)
\fmfleft{i}
\fmfright{o}
\fmf{wiggly}{i,v1}
\fmf{plain,left,tension=.3}{v1,v2}
\fmf{plain,right,tension=.3}{v1,v2}
\fmf{wiggly}{v2,o}
\end{fmfgraph}}\,
=
-\delta^{ab}g^4N\frac{\Gamma(2-\omega)
\Gamma(\omega)\Gamma(\omega-1)}{(4\pi)^{\omega} \Gamma(2\omega)}\cdot
(10-2\omega)\frac{ \delta_{\mu\nu}-p_\mu p_\nu/p^2}{p^{2-2\omega}}
\]
\item four flavors of four-component Majorana fermions in the adjoint
representation:
\[
\parbox{20mm}{
\begin{fmfgraph}(20,15)
\fmfleft{i}
\fmfright{o}
\fmf{wiggly}{i,v1}
\fmf{dashes,left,tension=.3}{v1,v2}
\fmf{dashes,right,tension=.3}{v1,v2}
\fmf{wiggly}{v2,o}
\end{fmfgraph}}\,
=
-\delta^{ab}g^4N\frac{\Gamma(2-\omega)
\Gamma(\omega)\Gamma(\omega-1)}{(4\pi)^{\omega} \Gamma(2\omega)}\cdot
16(\omega-1)\frac{ \delta_{\mu\nu}-p_\mu p_\nu/p^2}{p^{2-2\omega}}
\]
\end{itemize}
Note that these are the negative of the conventionally defined
self-energies.  Thus, to one loop order, the propagator for the
unrenormalized gluon, in Feynman gauge is
\begin{equation}
\Delta^{ab}_{\mu\nu}=g^2\delta^{ab} \frac{\delta_{\mu\nu}}{p^2}-g^4N
\frac{\Gamma(2-\omega) \Gamma(\omega)\Gamma(\omega-1)}{(4\pi)^{\omega}
\Gamma(2\omega)}\cdot 4(2\omega-1)\delta^{ab}\frac{
\delta_{\mu\nu}-p_\mu p_\nu/p^2}{p^{6-2\omega}}.
\label{gluon}
\end{equation}

Similarly, we can compute the one loop correction to the scalar propagator. 
It obtains corrections from:
\begin{itemize}
\item the scalar-vector intermediate state:
\[
\parbox{20mm}{
\begin{fmfgraph}(20,15)
\fmfleft{i}
\fmfright{o}
\fmf{plain}{i,v1}
\fmf{wiggly,left,tension=.2}{v1,v2}
\fmf{plain,tension=.2}{v1,v2}
\fmf{plain}{v2,o}
\end{fmfgraph}}\,
=
\delta^{ab}g^4N
\frac{\Gamma(2-\omega)
\Gamma(\omega)\Gamma(\omega-1)}{(4\pi)^{\omega} \Gamma(2\omega)}\cdot
4(2\omega-1)\frac{ \delta_{ij}}{p^{6-2\omega}}
\]
\item and the fermion loop:
\[
\parbox{20mm}{
\begin{fmfgraph}(20,15)
\fmfleft{i}
\fmfright{o}
\fmf{plain}{i,v1}
\fmf{dashes,left,tension=.3}{v1,v2}
\fmf{dashes,right,tension=.3}{v1,v2}
\fmf{plain}{v2,o}
\end{fmfgraph}}\,
=
-\delta^{ab}g^4N
\frac{\Gamma(2-\omega)
\Gamma(\omega)\Gamma(\omega-1)}{(4\pi)^{\omega} \Gamma(2\omega)}\cdot
8(2\omega-1)\frac{ \delta_{ij}}{p^{6-2\omega}}
\]
\end{itemize}
Thus, to one loop order, the (unrenormalized) scalar propagator is
\begin{equation}
D^{ab}_{ij}=g^2\delta^{ab}
\frac{\delta_{ij}}{p^2}-g^4N
\frac{\Gamma(2-\omega)
\Gamma(\omega)\Gamma(\omega-1)}{(4\pi)^{\omega} \Gamma(2\omega)}\cdot
4(2\omega-1)\frac{ \delta_{ij}\delta^{ab}}{p^{6-2\omega}}.
\label{scalar}
\end{equation}
Note that, aside from vector indices, the scalar and vector
propagators are identical. Also, note that the self-energy corrections
have poles at $\omega=2$ which arise from an ultraviolet divergence.
If we were to compute correlators of local renormalized fields, it
would be necessary to add a counterterm to the action in order to
cancel these ultraviolet singularities. Here, for purpose of computing
the Wilson loop, we leave them unrenormalized.

\subsection{Ladder diagrams}

The ladder-like diagrams to order $g^4N^2$ contribute
\[
\Sigma_1=\frac{g^4N^2}{6}\oint_{\tau_1>\tau_2>\tau_3>\tau_4}
d\tau_1\,d\tau_2\,d\tau_3\,d\tau_4\,\frac{(\lvert{\dot
    x}^{(1)}\rvert\lvert{\dot x}^{(2)}\rvert -{\dot x}^{(1)}\cdot{\dot
    x}^{(2)}) (\lvert{\dot x}^{(3)}\rvert\lvert{\dot x}^{(4)}\rvert
  -{\dot x}^{(3)}\cdot{\dot x}^{(4)})} {(\lvert
  x^{(1)}-x^{(2)}\rvert^2\lvert x^{(3)}-x^{(4)}\rvert^2)^{\omega-1}}.
\]
This result is finite when $x(\tau)$ is a smooth curve. 

\subsection{Insertion of one loop corrections to propagators}

Using the one loop self-energies of the vector and scalar fields that
we found in section 2.2, we can find the corrections to the
expression (\ref{leadloop}) resulting from insertion of one loop into
the vector and scalar field propagators.

We begin with the correction to the scalar propagator in momentum
space.  From equation (\ref{scalar}) it is
\[
-\delta^{ab}g^4N\frac{ \Gamma(2-\omega)\Gamma(\omega)\Gamma(\omega-1)}{(4\pi)^\omega
\Gamma(2\omega)}
4(2\omega-1)
\frac{1}{[p^2]^{3-\omega} }.
\]
The Fourier transform of this expression, computed using (\ref{ft}), is
\[
-\delta^{ab}g^4N\frac{\Gamma^2(\omega-1)} 
{2^{5}\pi^{2\omega}(2-\omega)(2\omega-3)}
\frac{1}{[x^2]^{2\omega-3} }.
\]
If we now combine this with the analogous expression for the vector
field propagator and compute the correction to the circular loop the
result is
\begin{equation}
\Sigma_2=-g^4 N^2\frac{\Gamma^2(\omega-1)} 
{2^{7}\pi^{2\omega}(2-\omega)(2\omega-3)}
\oint d\tau_1\, d\tau_2 \frac{ \left|{\dot x}^{(1)}\right|
\left| {\dot x}^{(2)}\right|-{\dot x}^{(1)}\cdot{\dot x}^{(2)}
}{\bigl[(x^{(1)}-x^{(2)})^2\bigr]^{2\omega-3} },
\label{result2}
\end{equation}
where a factor of $N^2/2$ came from taking the trace over gauge group
generators, a factor of $1/N$ came from the
normalization of the Wilson loop, and an additional factor of $1/2$ came
from the combinatorics of expanding the Wilson loop operator to second
order.  We see that the integrand is identical to (\ref{leadloop})
with a correction to the coefficient
\[
\frac{g^2N}{4\pi^2}\mapsto \frac{g^2N}{4\pi^2}-\frac{g^4N^2\Gamma^2(\omega-1)}
{128\,\pi^{2\omega}(2-\omega)(2\omega-3)}.
\]
The coefficient diverges in four dimensions (at $\omega=2$).

\subsection{Diagrams with one internal vertex}

The order $g^4N^2$ contributions with one internal vertex come when we
Taylor expand $W(C)$ to third order in $A$ and $\Phi$ and Wick
contract it with the relevant vertices to obtain the quantities:
\begin{gather*}
\frac{i^3}{3!}\int d\tau_1\,d\tau_2\,d\tau_3\Bigl<\Tr\path[A(\tau_1)A(\tau_2)A(\tau_3)]
\Bigl(-\int d^4y f^{abc} \partial_\mu \phi_i^a(y)A^b_\mu(y)\phi^c_i(y)\Bigr)\Bigr>,\\
\frac{i}{2!1!}\int d\tau_1\,d\tau_2\,d\tau_3\Bigl<\Tr\path[\Phi(\tau_1)A(\tau_2)\Phi(\tau_3)]
\Bigl(-\int d^4y f^{abc}\partial_\mu A_\nu^a(y)A_\mu^b(y)A_\nu^c(y)\Bigr)\Bigr>,
\end{gather*}
where $A(\tau)=A^a_\mu(x){\dot x}^\mu(\tau)T^a$ and
$\Phi(\tau)=\Phi^a(x)\lvert{\dot x}\rvert T^a$.  The minus signs in
both vertices come from the expansion of $e^{-S}$ (we work in
Euclidean space). The sum of the two diagrams is
\begin{equation}
\begin{split}
\Sigma_3&=
-\frac{g^4N^2}{4}\oint d\tau_1\,d\tau_2\,d\tau_3\,\epsilon(\tau_1\,\tau_2\,\tau_3)
(\lvert{\dot x}^{(1)}\rvert\lvert{\dot x}^{(3)}\rvert
-{\dot x}^{(1)}\cdot{\dot x}^{(3)})
\\&\qquad
\times {\dot x}^{(2)}\cdot\frac{\partial}{\partial x^{(1)}}
\int d^{2\omega}w\,\Delta(x^{(1)}-w)\Delta(x^{(2)}-w)\Delta(x^{(3)}-w).
\end{split}
\label{3part}
\end{equation}
Here, $\epsilon$ is the antisymmetric path ordering symbol: we define
$\epsilon(\tau_1\,\tau_2\,\tau_3)=1$ for $\tau_1>\tau_2>\tau_3$ and
let $\epsilon$ be antisymmetric under any transposition of
$\tau_i$. It is straightforward to introduce Feynman parameters and do
the integral over $w$ to obtain
\begin{equation}
\begin{split}
\Sigma_3&=
g^4N^2
\frac{\Gamma(2\omega-2)}
{2^{7}\pi^{2\omega}}
\int_0^1 d\alpha\,d\beta\,d\gamma\,(\alpha\beta\gamma)^{\omega-2}
\delta(1-\alpha-\beta-\gamma)
\oint d\tau_1\,d\tau_2\,d\tau_3\,\epsilon(\tau_1\,\tau_2\,\tau_3)
\\&\qquad\times
\frac{\bigl(\lvert{\dot x}^{(1)}\rvert\lvert{\dot x}^{(3)}\rvert-{\dot x}^{(1)}\cdot{\dot x}^{(3)}\bigr) 
\bigl(\alpha(1-\alpha){\dot x}^{(2)}\cdot x^{(1)}-\alpha\gamma{\dot x}^{(2)}\cdot x^{(3)} -\alpha\beta{\dot x}^{(2)}\cdot x^{(2)}\bigr)
}{
\bigl[\alpha\beta\lvert x^{(1)}-x^{(2)}\rvert^2 +
\alpha\gamma\lvert x^{(1)}-x^{(3)}\rvert^2 +
\beta\gamma\lvert x^{(3)}-x^{(2)}\rvert^2\bigr]^{2\omega-2}  
}.
\label{mastereqn}
\end{split}
\end{equation}

\section{Cancellation of divergences to order $g^4N^2$}\label{canl}

A logarithmic divergence arises in the integral (\ref{3part}) from
where $\tau_1$ is coincident with $\tau_2$.  This divergence should
cancel with the divergence in the coefficient of $\Sigma_2$ in
(\ref{result2}) so that the order $g^4N^2$ contribution
\[
\Sigma_1+\Sigma_2+\Sigma_3
\]
is finite.
In extracting the divergences from (\ref{3part}),
we must consider the integral
\begin{equation}
G(\tau_i)=
\int d^4w\,\Delta(w-x^{(1)})\Delta(w-x^{(2)})\Delta(w-x^{(3)}),
\label{A1}
\end{equation}
in detail. This integral is singular in the limit $x^{(2)}\to
x^{(1)}$.  The divergent contribution comes from the integration over
$w$ close to $x^{(1)}$ and $x^{(2)}$. We can approximate
$|w-x^{(3)}|\approx |x^{(1)}-x^{(3)}|$ introducing simultaneously an
infrared cutoff $\delta$, so that the divergent part of (\ref{A1})
takes the form
\begin{equation}
G(\tau_i)\sim
\Delta(x^{(1)}-x^{(3)})\int^\delta d^4w\, \Delta(w-x^{(1)})\Delta(w-x^{(2)}).
\label{A2}
\end{equation}
As follows from dimensional counting, this integral depends only on
the dimensionless ratio $|x^{(1)}-x^{(2)}|/\delta$. The limit where
$x^{(1)}$ and $x^{(2)}$ are coincident is then equivalent to the limit
of infinite $\delta$.  Without the infrared cutoff the integral would
diverge logarithmically, so up to terms regular in the limit
$\delta\rightarrow\infty$, we get
\begin{equation}\label{A3}
G(\tau_i)\sim
\frac{1}{64\pi^6}\frac{1}{\lvert x^{(1)}-x^{(3)}\rvert^2}
\int \frac{d^4w}{w^4}=
-\frac{1}{64\pi^4}
\frac{\log\,\lvert x^{(1)}-x^{(2)}\rvert^2/\delta^2}{|x^{(1)}-x^{(3)}|^2}.
\end{equation}
Since $G$ only goes as a logarithm as points approach each other,
(\ref{3part}) receives divergent contributions only from $x^{(1)}$
near $x^{(2)}$; the parts of the integral with $x^{(1)}$ or $x^{(2)}$
near $x^{(3)}$ are finite. This divergence is regularized by cutting
off the integral over $\tau_1$. Since the overall divergence is
logarithmic, the result is independent of the method of
regularization. Writing $\tau=\tau^{(1)}-\tau^{(2)}$ and Taylor
expanding $x^{(1)}=x^{(2)}+\dot{x}^{(2)}\tau +\cdots$, we see that
the divergent part of (\ref{3part}) is
\[
\begin{split}
\Sigma_3&\sim
-\frac{g^4N^2}{128\pi^4}\oint d\tau_2\oint d\tau_3 \frac{
\lvert{\dot x}^{(2)}\rvert\lvert{\dot x}^{(3)}\rvert
-{\dot x}^{(2)} \cdot{\dot x}^{(3)}}
{\lvert x^{(2)}-x^{(3)}\rvert^2}\int d\tau\,\sign\tau
\frac{{\dot x}^{(2)}\cdot(x^{(1)}-x^{(2)})}
{\lvert x^{(1)}-x^{(2)}\rvert^2}
\\&=
-\frac{g^4N^2}{64\pi^4}\oint d\tau_2\oint d\tau_3 \frac{
\lvert{\dot x}^{(2)}\rvert\lvert{\dot x}^{(3)}\rvert
-{\dot x}^{(2)} \cdot{\dot x}^{(3)}}
{\lvert x^{(2)}-x^{(3)}\rvert^2}\log\epsilon.
\end{split}
\]
This cancels exactly against (\ref{result2}), for one should replace
the pole $1/(2-\omega)$ at $\omega=2$ by $-2\log\epsilon$.

\section{The circular loop}

In this section, we consider a circular Wilson loop, whose radius we
can assume to be unity. A convenient parameterization of this loop is
\begin{equation}
x(\tau)=(\cos \tau, \sin \tau, 0, 0).
\label{circle}
\end{equation}

\subsection{Summing the planar ladder graphs}

First, we will see how to sum all planar diagrams which have no
internal vertices.  These include all ladder and rainbow diagrams.
Our strategy is the following: the large $N$ limit with $g^2N$
held fixed is given by planar diagrams.  It is well known that each
planar graph will contain the same group theoretical factors
\cite{tHo82}.  We observe that, in fact, each planar diagram without
internal vertices gives an identical contribution to the loop
expectation value.  We choose a fixed ordering of the times and
compute a particular, convenient diagram.  Then we multiply by
the number of diagrams that occur to that order and sum over
all orders.

First, consider the $2n$-th order term in the Taylor expansion of the loop
\[
\frac{1}{N}
\int_0^{2\pi} d\tau_1\int_0^{\tau_1}d\tau_2\cdots 
\int_0^{\tau_{2n-1}}d\tau_{2n}  \Tr
\bigl<\bigl(iA(\tau_1)+\Phi(\tau_1)\bigr) 
\cdots
\bigl(iA(t_{2n})+\Phi(t_{2n})\bigr)\bigr>.
\]
Here we have chosen a particular time ordering, which cancels the
factor of $1/(2n)!$ which would come from the Taylor expansion of the
exponential.  We are interested in all Wick contractions which
represent planar diagrams. Note that for the circular loop
\begin{equation}\label{contr}
\langle
\bigl(iA^{a}(\tau_1)+\Phi^{a}(\tau_1)\bigr) 
\bigl(iA^{b}(\tau_2)+\Phi^{b}(\tau_2)\bigr) 
\rangle_0
=\frac{g^2\delta^{ab}}{4\pi^2}
\frac{\lvert{\dot x}^{(1)}\rvert\lvert{\dot x}^{(2)}\rvert-{\dot x}^{(1)}
\cdot {\dot x}^{(2)}}{\lvert x^{(1)}-x^{(2)}\rvert^2}
=\frac{g^2\delta^{ab}}{8\pi^2}.
\end{equation}
Thus, the contributions of all (free-field) Wick contractions giving
planar diagrams are identical (if we sum over
all ways of choosing scalar or gluon lines). The color factor can be
computed by repeated application of the identity:
\[
T^aT^a=\tfrac{N}{2}\,{\bf 1}.
\]
Thus, for the sum of ladder-like diagrams with $n$ propagators we obtain 
\begin{equation}
\frac{(g^2N/4)^n}{(2n)!}
\times(\text{\# of planar graphs with }n\text{ internal lines}),
\label{contrib}
\end{equation}
where the factor $1/(2n)!$ accounts for the integral over the $\tau^{(i)}$.

We now have the task of counting the number of planar graphs with $n$
internal lines.  Any such diagram with $n+1$ propagators can be
uniquely decomposed as
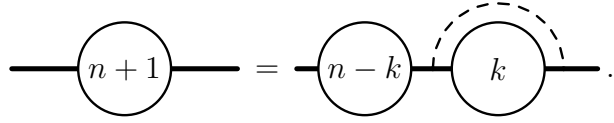
\[
\parbox{30mm}{
\begin{fmfgraph*}(30,15)
\fmfleft{l}
\fmfright{r}
\fmf{plain,width=2}{l,v,r}
\fmfv{d.sh=circle,d.f=empty,d.si=35,l=$n+1$,l.d=0}{v}
\end{fmfgraph*}}
\;=\;
\parbox{40mm}{
\begin{fmfgraph*}(40,15)
\fmfleft{l}
\fmfright{r}
\fmf{plain,tension=50,width=2}{l,b1}
\fmf{plain,tension=50,width=2}{b1,v1}
\fmf{plain,tension=50,width=2}{v1,b2}
\fmf{plain,tension=50,width=2}{b2,v2}
\fmf{plain,tension=100,width=2}{v2,r}
\fmf{dashes,left}{v1,v2}
\fmfv{d.sh=circle,d.f=empty,d.si=35,l=$n-k$,l.d=0}{b1}
\fmfv{d.sh=circle,d.f=empty,d.si=35,l=$k$,l.d=0}{b2}
\end{fmfgraph*}}\;.
\]
If we define $A_{n+1}$ as the number of such diagrams then $A_{n+1}$
satisfies the recursion relation
\[
A_{n+1}=\sum_{k=0}^nA_{n-k}A_k,
\]
with $A_0=1$. If we define a generating function $f$ by
\[
f(z)=\sum_{n=0}^\infty A_nz^n,
\] 
then $f$ satisfies $zf^2(z)=f(z)-1$.
So,
\begin{equation}
f(z)=\frac{1-\sqrt{1-4z}}{2z}=\sum_{n=0}^\infty \frac{(2n)!}{(n+1)!n!}z^n.
\label{recurcount}
\end{equation}
The sign of the square root is chosen by requiring that $f$ be finite
at $z=0$. Hence
\begin{equation}
A_n=\frac{(2n)!}{(n+1)!n!},
\label{count}
\end{equation}
so the sum of \textit{all} planar diagrams without vertices on the
loop is, from (\ref{contrib}) and (\ref{count})
\begin{equation}
\left< W(C)\right>_{\rm ladders}
=\sum_{n=0}^\infty \frac{(g^2N/4)^n}{(n+1)!n!}=
-\frac{2}{\sqrt{g^2N}}I_1(\sqrt{g^2N}).
\label{circladder}
\end{equation}
Thus, the large
$g^2N$ behavior is
\begin{equation}
\left< W(C)\right>_{\rm ladders}
\sim \frac{e^{\sqrt{g^2N}}}{(\pi/2)^{1/2}(g^2N)^{3/4}}.
\label{circpert}
\end{equation}
The supergravity prediction is that
\begin{equation}
\langle W(C)\rangle_{\text{AdS/CFT}} \sim e^{\sqrt{g^2N}},
\label{circads}
\end{equation}
so the ladder diagrams have the same leading behavior as the
prediction of the AdS/CFT correspondence.

It is worth mentioning that the cancellation of coordinate dependence
in the Wick contraction (\ref{contr}) maps the problem of summing the
ladder-like diagrams to the zero dimensional theory. In particular,
the number of planar graphs with no vertices and $n$ propagators can
be calculated from the infinite $N$ limit of the matrix integral
\cite{Bre78}
\[
A_n=\Bigl<\frac{1}{N}\Tr M^{2n}\Bigr>=
\frac{1}{Z}\,{\int[dM]\,\frac{1}{N}\Tr M^{2n}\exp
\Bigl\{-\frac{N}{2}\Tr M^2\Bigr\}},
\]
where
\[
Z=\int[dM]\, \exp\Bigl\{-\frac{N}{2}\Tr M^2\Bigr\}.
\]
This can be evaluated by extracting the term of order $z^{2n}$ in the
Taylor expansion of the zero-dimensional Wilson loop
\cite{Mak91}:
\begin{equation}
\Omega(z)=\left\langle\frac{1}{N}\Tr \frac{1}{1-zM}\right\rangle.
\end{equation}
Using the identity
\[
\frac{1}{(2n)!}=\oint_{\cal C} \frac{dz}{2\pi i}\frac{e^z}{z^{2n+1}}
\]
for a positively oriented contour $\cal C$ containing the origin, 
we can represent the sum of the ladder diagrams as
\begin{equation}\label{contouri}
\langle W(C)\rangle_{\rm ladders}
=\oint_{\cal C} \frac{dz}{2\pi i}\frac{e^z}{z}\Omega(g^2N/4z),
\end{equation}
where $\cal C$ must be chosen large enough to encircle all
singularities of the integrand. The function $\Omega(z)$ satisfies a
zero-dimensional loop equation \cite{Mak91} that follows from the
identity
\[
0=\int[dM]\frac{\partial}{\partial M_{ij}}
\left\{\Bigl(\frac{1}{1-zM}\Bigr)_{ij}\exp(-\tfrac{N}{2}\Tr M^2)\right\}
\]
and large $N$ factorization. In the infinite $N$ limit,
this equality reduces to the algebraic equation for $\Omega(z)$:
\[
z\Omega^2(z)-\frac{1}{z}\Omega(z)+\frac{1}{z}=0.
\]
This has solution
\[
\Omega(z)=\frac{1-\sqrt{1-4z^2}}{2z^2}.
\]
Substituting $z^2\to z$, this is the same as (\ref{recurcount}).
Shrinking the contour of integration in (\ref{contouri}) to the branch
cut of $\Omega(z)$, we obtain
\[
\left< W(C)\right>_{\rm ladders}
=4\int_{-1}^{1}\frac{dx}{2\pi}e^{\sqrt{g^2N}x}\sqrt{1-x^2},
%=-\frac{2i}{\sqrt{g^2N}}J_1(i\sqrt{g^2N}),
\]
which is just the Bessel function (\ref{circladder}). 

\subsection{Diagrams with vertices}

The sum of the diagrams with one internal vertex attaching to three
points on the Wilson loop is given by the expression
(\ref{mastereqn}). It is convenient to abbreviate
$\tau_{ij}=\tau_i-\tau_j$. For a circular loop, $\lvert
x^{(i)}\rvert^2=1$, $\lvert
x^{(i)}-x^{(j)}\rvert^2=2(1-\cos\tau_{ij})$,
$x^{(i)}\cdot\dot{x}^{(j)}=\sin\tau_{ij}$, and $\dot x^{(i)}\cdot\dot
x^{(j)}=\cos\tau_{ij}$. Thus, from (\ref{3part})
\begin{equation}
\begin{split}
\Sigma_3&=g^4N^2
\frac{\Gamma(2\omega-2)}{2^{2\omega+5}\pi^{2\omega}}
\int_0^1 d\alpha\,d\beta\,d\gamma\,(\alpha\beta\gamma)^{\omega-2}\delta(1-\alpha-\beta-\gamma)
\\&
\qquad\times\oint d\tau_1\,d\tau_2\,d\tau_3\,
\frac{\epsilon(\tau_1\,\tau_2\,\tau_3) (1-\cos\tau_{13})
\bigl(\alpha(1-\alpha)\sin\tau_{12}+
\alpha\gamma\sin\tau_{23}\bigr)
}
{
\bigl[\alpha\beta(1-\cos\tau_{12})+
\beta\gamma(1-\cos\tau_{23})+
\gamma\alpha(1-\cos\tau_{13})
\bigr]^{2\omega-2}  
}.
\end{split}
\label{master}
\end{equation}
We are going to use integration by parts to rewrite (\ref{master})
as a sum of a term which will cancel with the order $g^4N^2$ diagrams
with internal vertices in (\ref{result2}) and a term
which vanishes when $\omega=2$. For compactness, write the denominator
\[
\Delta=\alpha\beta(1-\cos\tau_{12})+
\beta\gamma(1-\cos\tau_{23})+
\gamma\alpha(1-\cos\tau_{13}).
\]

Consider the identity
\begin{equation}
\oint d\tau_1\,d\tau_2\,d\tau_3\,\frac{\partial}{\partial \tau_1}
\frac{\epsilon(\tau_1\,\tau_2\,\tau_3)
(1-\cos\tau_{13})
}
{
\Delta^{2\omega-3}  
}
=0.
\end{equation}
Using
\begin{equation}
\frac{\partial}{\partial \tau_1}\epsilon(\tau_1\,\tau_2\,\tau_3)=2\delta(\tau_{12})-2\delta(\tau_{13}),
\label{epsilderiv}
\end{equation}
that $\alpha+\beta+\gamma=1$, and the fact that the integrand in vanishes when $\tau_1=\tau_3$, we get
%\[
%\oint d\tau_1\,d\tau_2\,d\tau_3 \left\{2\delta(\tau_{12})\frac{1}{[\gamma(1-\gamma)]^{2\omega-3}}
%\frac{1}{ [1-\cos\tau_{13}]^{2\omega-4}}
%+\epsilon(\tau_1\,\tau_2\,\tau_3)
%\frac{\partial}{\partial \tau_1}\frac{ (1-\cos\tau_{13})
%}
%{
%\Delta^{2\omega-3}  
%}\right\}=0
%\]
%which can be rearranged to read
\begin{multline}
\oint d\tau_1\,d\tau_2\,d\tau_3
\biggl\{-
\frac{\sin\tau_{13}
\bigl(\alpha\beta(1-\cos\tau_{12})
+\beta\gamma(1-\cos\tau_{23})
+\gamma\alpha(1-\cos\tau_{13})\bigr)
}
{
\Delta^{2\omega-2}
}
\\
+
(2\omega-3)
\frac{
(1-\cos\tau_{13})(\alpha\beta\sin\tau_{12}+\gamma\alpha\sin\tau_{13})
}
{
\Delta^{2\omega-2}
}
\biggr\}\epsilon(\tau_1\,\tau_2\,\tau_3)\\
=2\oint d\tau_1\,d\tau_2\,\frac{1}{[\gamma(1-\gamma)]^{2\omega-3}}\frac{1}{[1-\cos\tau_{12}]^{2\omega-4}}.
\end{multline}
Now, change variables in the first term on the left hand side so that
the cosines in the denominator all have argument $\tau_{13}$ (note
that this involves permuting $\alpha$, $\beta$, and $\gamma$ as well,
so the following holds only after inserting
$\delta(1-\alpha-\beta-\gamma)$ and integrating these parameters over
the unit cube). Then add part of the second term, so that the
remaining part is proportional to $(2\omega-4)$ to obtain
\begin{multline*}
\oint d\tau_1\,d\tau_2\,d\tau_3\,\epsilon(\tau_1\,\tau_2\,\tau_3)
\biggl\{
\frac{(1-\cos\tau_{13})
\bigl(
\alpha(1-\alpha)\sin\tau_{12}
+\alpha\gamma\sin\tau_{23})
}
{
\Delta^{2\omega-2}
}
\\
+
(2\omega-4)
\frac{
(1-\cos\tau_{13})(\alpha\beta\sin\tau_{12}+\gamma\alpha\sin\tau_{13})
}
{
\Delta^{2\omega-2}
}
\biggr\}\\
=2\oint d\tau_1\,d\tau_2\,\frac{1}{[\gamma(1-\gamma)]^{2\omega-3}}\frac{1}{[1-\cos\tau_{12}]^{2\omega-4}}.
\end{multline*}
The first term on the left hand side is precisely the term occurring in
(\ref{master}). Note that the second term can be rewritten as
\begin{multline*}
-\frac{2\omega-4}{2\omega-3}
\oint d\tau_1\,d\tau_2\,d\tau_3\,\epsilon(\tau_1\,\tau_2\,\tau_3) (1-\cos\tau_{13})\frac{\partial}{\partial \tau_1}
\frac{1}
{\Delta^{2\omega-3}}
 \\
=\frac{2\omega-4}{2\omega-3}
\biggl\{\oint d\tau_1\,d\tau_2\,d\tau_3\,\epsilon(\tau_1\,\tau_2\,\tau_3)
\frac{\sin\tau_{13}}
{\Delta^{2\omega-3}}\\
+2\oint d\tau_1\,d\tau_2\,\frac{1}{[\gamma(1-\gamma)]^{2\omega-3}}\frac{1}{[1-\cos\tau_{12}]^{2\omega-4}}\biggr\},
\end{multline*}
using integration by parts. Finally, we have
\begin{multline*}
\oint d\tau_1\,d\tau_2\,d\tau_3\,\epsilon(\tau_1\,\tau_2\,\tau_3)
\frac{(1-\cos\tau_{13})
\bigl(
\alpha(1-\alpha)\sin\tau_{12}
+\alpha\gamma\sin\tau_{23})
}{\Delta^{2\omega-2}}
\\
=-
\frac{2\omega-4}{2\omega-3}
\oint d\tau_1\,d\tau_2\,d\tau_3\,\epsilon(\tau_1\,\tau_2\,\tau_3)
\frac{\sin\tau_{13}}
{\Delta^{2\omega-3}}\\
+\frac{2}{2\omega-3}\oint d\tau_1\,d\tau_2\,\frac{1}{[\gamma(1-\gamma)]^{2\omega-3}}\frac{1}{[1-\cos\tau_{12}]^{2\omega-4}}.
\end{multline*}
If we symmeterize, the integral in the first term on the right hand
side may be rewritten (setting $\omega=2$) as
\[
\frac{1}{3}\frac{2\omega-4}{2\omega-3}
\oint d\tau_1\,d\tau_2\,d\tau_3\,\epsilon(\tau_1\,\tau_2\,\tau_3)
\frac{
\sin\tau_{12}
+\sin\tau_{23}
+\sin\tau_{31}
}
{\alpha\beta(1-\cos\tau_{12})+
\beta\gamma(1-\cos\tau_{23})+
\gamma\alpha(1-\cos\tau_{13})}.
\]
This integration is completely finite. Since this term appears with a
coefficient that vanishes when $\omega=2$, it vanishes in four
dimensions.  Inserting this expression into (\ref{master}), we have
the result
\begin{equation}
\Sigma_3=
g^4N^2
\frac{ \Gamma^2(\omega-1)} {2^{2\omega+4}\pi^{2\omega}(2\omega-3)(2-\omega)}
\oint d\tau_1\,d\tau_2\,
\frac{1}{ [1-\cos\tau_{12}]^{2\omega-4} }+{\cal O}(2\omega-4).
\label{result1}
\end{equation}
If we specialize (\ref{result2}) to the case of a circular loop, we
obtain
\begin{equation}
\Sigma_2=-g^4N^2\frac{\Gamma^2(\omega-1)} 
{2^{2\omega+4}\pi^{2\omega}(2-\omega)(2\omega-3)}
\oint d\tau_1\,d\tau_2\, \frac{1}{[1-\cos\tau_{12}]^{2\omega-3} }.
\label{result3}
\end{equation}
The two contributions (\ref{result1}) and (\ref{result3})
cancel exactly when $2\omega=4$:
\[
\Sigma_2+\Sigma_3=0.  
\]

\section{Anti-Parallel lines}

\subsection{Summing Ladder Diagrams}

It is possible to sum the planar ladder diagrams for antiparallel
lines separated by a distance $L$ \cite{Eri99}.  The calculation can
be done by considering the more general case of a sum $\Gamma(S,T)$ of
ladder diagrams for two finite antiparallel lines of lengths $S$ and
$T$ (we take $S,T\gg L$). The result we are interested in arises in
studying the large $S=T$ behavior. The sum of ladder diagrams
satisfies the recursion relation
\begin{equation}
\Gamma(S,T)=1+\int_0^Sds\int_0^Tdt\,
\Gamma(s,t)\frac{g^2N}{4\pi^2 [(s-t)^2+L^2]},
\label{inteq}
\end{equation}
where the second factor in the integrand is the sum of 
vector and scalar propagators. $\Gamma(S,T)$ satisfies the
boundary conditions
\begin{equation}
\Gamma(S,0)=\Gamma(0,T)=1.
\label{bc}
\end{equation}
Taking derivatives of (\ref{inteq}) we obtain  the differential equation
\begin{equation}
\frac{\partial ^2\Gamma(S,T)}{\partial S\,\partial T}=
\frac{g^2N}{4\pi^2\left[ (S-T)^2+L^2\right]}\Gamma(S,T).
\end{equation}
This equation is separable in the variables $x=(S-T)/L$ and $y=(S+T)/L$.  (However,
as we shall see in the following,
the boundary conditions are not conveniently expressed in terms of these variables.)
The separated ansatz is 
\begin{equation} 
\Gamma[x,y]=\sum_n c_n \psi_n(x)\exp(\Omega_n y/2),
\label{sep}
\end{equation}
where $c_n$ are constants which must be determined so that the 
boundary condition (\ref{bc}) is satisfied and 
$\psi_n(x)$ are solutions of the Schr\"odinger equation
\begin{equation}
\left[-\frac{d^2}{dx^2}-\frac{g^2N}{4\pi^2(x^2+1)}
\right]\psi_n(x)=
-\frac{\Omega_n^2}{4}\psi_n(x).
\label{schro}
\end{equation}
The largest $\Omega_n$ with non-vanishing contribution in (\ref{sep})
dominates the large $y$ for fixed $x$ (large $S$ and $T$) asymptotics
of $\Gamma$.  In order to see that the lowest bound state of the
Schr\"odinger operator in (\ref{schro}) contributes, we consider the
Laplace transform of $\Gamma[x,y]$. Accounting for the boundary
condition (\ref{bc}) which implies that $\Gamma[x,y]$ vanishes when
$y=\lvert x\rvert$, we see that the Laplace transform is the
resolvent of the Schr\"odinger equation,
\[
\int_{\lvert x\rvert}^\infty dy\,e^{-py}\Gamma(x,y)=2\sum_n
\frac{\bar{\psi}_n(0)\psi_n(x)}{p^2-\Omega^2_n/4}.
\]
Since the potential is symmetric and the ground state has no nodes
($\psi_0(0)\neq 0$), the ground state eigenvalue gives the largest
$\Omega_n$ contributing to (\ref{sep}).

It is possible to find the ground state eigenvalues of (\ref{schro})
for small and large $g^2N$. We find
\begin{equation}\label{glarge}
\ln\Gamma(T,T)=
\left(g^2N/4\pi
-\frac{g^4N^2}{8\pi^3}\ln\frac{1}{g^2N}+\cdots\right)\frac{T}{L},  
\end{equation}
for $g^2N\ll 1$ and
\begin{equation}\label{gsmall}
\ln\Gamma(T,T)=
\left(\sqrt{g^2N}/\pi-1+{\cal O}(1/\sqrt{g^2N})\right)\frac{T}{L}, 
\end{equation}
for $g^2N\gg 1$. 
%In the large 't~Hooft coupling limit, the AdS/CFT
%computation gives
%\[
%\langle W\rangle\sim\exp\Bigl(\frac{4\pi^2\sqrt{2g^2N}}{\Gamma^4(1/4)}T\Bigr),
%\]
%for $g^2N\gg 1$. The coefficients in the two calculations are similar, with
%\[
%1/\pi\approx 0.318\qquad\text{and}
%\qquad 4\pi^2\sqrt{2}/\Gamma^4(1/4)\approx 0.323.
%\]

\subsection{Diagrams with vertices}

The sum of the diagrams with one internal vertex and three lines going
to the Wilson loop is given by (\ref{mastereqn}).  For anti-parallel
lines, $\tau_1$ and $\tau_3$ must be on opposite lines so that the
factor $\left|{\dot x}^{(1)}\right| \left|{\dot x}^{(3)}\right|-{\dot
  x}^{(1)}\cdot {\dot x}^{(3)}$ is non-zero.  In that case, it
provides a factor of 2.  Furthermore, we will find a non-vanishing
result only when $\tau_2$ is on the same line as $\tau_1$.  There are two
possible configurations like this, which provides a further factor of
2.  Taking the parameterization of the lines to be
\begin{gather}
x^{(1)}=\left( \tau_1, L/2,0,0\right) \nonumber \\
x^{(2)}=\left( \tau_2, L/2,0,0\right) \nonumber \\
x^{(3)}=\left( -\tau_3, -L/2,0,0\right) \nonumber
\end{gather}
we obtain
\[
\begin{split}
\Sigma_3&=g^4N^2 \frac{\Gamma(2\omega-2) }{2^5 \pi^{2\omega} }
\int_0^1 d\alpha\,d\beta\,d\gamma\,\left(\alpha\beta\gamma\right)^{\omega-2}
\delta(1-\alpha-\beta-\gamma)\int d\tau_1\,d\tau_2\,d\tau_3\,\epsilon(\tau_1\,\tau_2\,\tau_3)
\\&\qquad\qquad
\times \frac{[ \alpha\beta(\tau_1-\tau_2)+
\gamma\alpha(\tau_1+\tau_3)] } {
\bigl[ \alpha\beta(\tau_1-\tau_2)^2+\beta\gamma\bigl((\tau_2+\tau_3)^2+L^2\bigl)^2
+\gamma\alpha\bigl((\tau_3+\tau_1)^2+L^2\bigr)^2\bigr]^{2\omega-2} },
\end{split}
\]
which can be written as
\[
\begin{split}
\Sigma_3&=-g^4N^2 \frac{\Gamma(2\omega-3) }{2^6 \pi^{2\omega} }
\int_0^1 d\alpha\,d\beta\,d\gamma\,\left(\alpha\beta\gamma\right)^{\omega-2}
\delta(1-\alpha-\beta-\gamma)\int d\tau_1\,d\tau_2\,d\tau_3\,\epsilon(\tau_1\,\tau_2\,\tau_3)
\\&\qquad\qquad
\times\frac{\partial}{\partial \tau_1} 
\frac{ 1 } {
\left[ \alpha\beta(\tau_1-\tau_2)^2+\beta\gamma\bigl((\tau_2+\tau_3)^2+L^2\bigr)
+\gamma\alpha\bigl( (\tau_3+\tau_1)^2+L^2\bigr)\right]^{2\omega-3} }.
\end{split}
\]
Then, using (\ref{epsilderiv}) we can do the integral over $\tau_1$ to obtain
\[
\begin{split}
\Sigma_3&=g^4N^2 \frac{\Gamma(2\omega-3) }{2^5 \pi^{2\omega} }
\int_0^1 d\alpha\,d\beta\,d\gamma\,
\frac{(\alpha\beta\gamma)^{\omega-2}\delta(1-\alpha-\beta-\gamma)} 
{\bigl[ \gamma(1-\gamma)\bigr]^{2\omega-3}}
\\&\qquad\qquad\times
\int_{-\infty}^\infty d\tau_2\,d\tau_3\,\frac{1}{\bigl[ (\tau_3+\tau_2)^2+L^2 \bigr]^{2\omega-3} },
\end{split}
\]
which yields
\[
\Sigma_3=g^4N^2 \frac{\Gamma^2(\omega-1)}
{32 \pi^{2\omega}(2\omega-3)(2-\omega)}
\int_{-\infty}^\infty d\tau_2\,d\tau_3\,\frac{1}{\bigl[ (\tau_3+\tau_2)^2+L^2\bigr]^{2\omega-3} }.
\]
This result is to be added to the diagrams which come from exchange of
a vector and scalar field, each with internal loop corrections.  The
sum of those two diagrams is
\begin{equation}
\Sigma_2=-g^4N^2\frac{\Gamma^2(\omega-1)}{32\pi^{2\omega}(2\omega-3)(2-\omega)}
\int d\tau_2\,d\tau_3\,\frac{ 1}{\bigl[ (\tau_3+\tau_2)^2+L^2\bigr]^{2\omega-3} },
\end{equation}
and, in the end,
\[
\Sigma_2+\Sigma_3=0.
\]
We see that the quantum corrections to the Wilson loop cancel 
identically in all spacetime
dimensions less than ten.

\section{Conclusions}

We have calculated contribution of planar diagrams without internal
vertices to Wilson loops in \N supersymmetric Yang-Mills theory. For the two particular
types of loops we have considered, circular and rectangular, these diagrams
exponentiate. This fact is rather unexpected, given that the
combinatorics of planar diagrams without vertices is different from
that of diagrams in the free $U(1)$ theory \cite{Cvi81,Cvi82,Gop94},
which exponentiate for obvious reasons. Because the ladder-like
diagrams exponentiate, the static potential is well defined in the
ladder approximation and is easy to extract from the expectation
values of the rectangular Wilson loop, (\ref{glarge}) and
(\ref{gsmall}) \cite{Eri99}. The dependence of the potential on the
coupling constant, when extrapolated to the strong coupling regime, is
very similar to, but not exactly the same as the one predicted by the
AdS/CFT correspondence.  We have no explanation for this similarity,
but the result for the circular loop indicates that this is not a mere
coincidence.

The planar diagrams without internal vertices for the circular Wilson
loop sum up into an expression for which the strong coupling limit
coincides with the prediction of the AdS/CFT correspondence.  We have
conjectured that diagrams with internal vertices cancel for this
Wilson loop in \N supersymmetric Yang-Mills theory. This conjecture is
supported both by a direct calculation at order $g^4N^2$ and by
agreement with the strong coupling results of the AdS/CFT
correspondence.  It seems that conformal symmetry plays a key role
here, as the conformal symmetry of the theory would na\"\i{}vely
suggest that all higher orders of perturbation theory should cancel.
While this is not the case, for reasons discussed above, it may be
true that the diagrams with internal vertices are constrained by
conformal symmetry to cancel amongst themselves. This is supported by
the fact that the remarkable cancellation occurs only in four
dimensions, the dimensionality in which conformal symmetry is present.

\section*{Acknowledgments}
This work is supported in part by NSERC of Canada.  G.W.S. is
supported by the the Ambrose Monell Foundation.  The work of K.Z. is
supported by PIMS. G.W.S. thanks Steve Adler, Ken Intriligator, Sasha
Polyakov and Frank Wilczek for helpful discussions.  J.K.E. and K.Z.
thank Yuri Makeenko for helpful discussions.

\appendix
\section{Useful formulae and notation}

The Euclidean space action of four dimensional \N supersymmetric Yang-Mills theory is
\begin{equation}
\begin{split}
S&=\int d^4x \frac{1}{2g^2}\Bigl\{ \frac{1}{2}\left(F_{\mu\nu}^a\right)^2
+\left(\partial_\mu \phi_i^a+f^{abc}A^b_\mu \phi_i^c\right)^2
+\bar\psi^a i\gamma^\mu\left(
\partial_\mu \psi^a+f^{abc}A_\mu^b\psi^c\right)
\\&\qquad
+if^{abc}\bar\psi^a\Gamma^i\phi^b_i\psi^c -\sum_{i<j}f^{abc}f^{ade}\phi_i^b\phi_j^c\phi_i^d\phi_j^e
+\partial_\mu\bar c^a\left(\partial_\mu c^a+f^{abc}A_\mu^b c^c\right)
+\xi( \partial_\mu A_\mu^a)^2\Bigr\}
\end{split}
\label{actionn}
\end{equation}
where 
\[
F_{\mu\nu}^a=\partial_\mu A_\nu^a -\partial_\nu A_\mu^a+
f^{abc}A^b_\mu A^c_\nu
\]
and $f^{abc}$ are the structure constants of the $U(N)$ Lie algebra,
\[
\left[ T^a,T^b\right]=if^{abc}T^c.
\]
The generators are normalized as
\[
\Tr T^aT^b=\frac{1}{2}\delta^{ab},
\]
and obey the identity
\[
\sum_{c,d}f^{acd}f^{bcd}=N\delta^{ab}.
\]

Now, $\psi^a$ is a sixteen component spinor obeying the Majorana
condition
\begin{equation}
\psi(x)=C\psi^*(x),
\end{equation}
where $C$ is the charge conjugation matrix.  $\Gamma^A=(\gamma^\mu,
\Gamma^i)$, for $\mu=1,...,4$ and $i=5,...,10$ are ten real $16\times
16$ Dirac matrices (in the 10-dimensional Majorana representation with
the Weyl constraint) obeying
\[
\Tr\left( \Gamma^A \Gamma^B \right) = 16 \delta^{AB}.
\]
We have chosen the covariant gauge fixing condition,
\begin{equation}
\partial_\mu A_\mu^a=0,
\end{equation}
and we work in Feynman gauge, where the gauge parameter is chosen
as $\xi=1$.  The appropriate action for ghost fields, $c^a(x)$ has
been included.

In Feynman gauge, the vector field propagator is
\[
\Delta^{ab}_{\mu\nu}(p)=
\parbox{10mm}
{\begin{fmfgraph}(10,0)
\fmfleft{i}
\fmfright{o}
\fmf{wiggly}{i,o}
\end{fmfgraph}}
=g^2\delta^{ab}\frac{\delta_{\mu\nu}}{p^2},
\]
the scalar propagator is
\[
D^{ab}_{ij}(p)=
\parbox{10mm}
{\begin{fmfgraph}(10,0)
\fmfleft{i}
\fmfright{o}
\fmf{plain}{i,o}
\end{fmfgraph}}
=g^2\delta^{ab}\frac{\delta_{ij}}{p^2},
\]
the fermion propagator is
\[
S^{ab}(p)=
\parbox{10mm}
{\begin{fmfgraph}(10,0)
\fmfleft{i}
\fmfright{o}
\fmf{dashes}{i,o}
\end{fmfgraph}}
=g^2\delta^{ab}\frac{-\gamma\cdot p}{p^2},
\]
and the ghost propagator is
\[
C^{ab}(p)=
\parbox{10mm}
{\begin{fmfgraph}(10,0)
\fmfleft{i}
\fmfright{o}
\fmf{dots}{i,o}
\end{fmfgraph}}
=g^2\delta^{ab}\frac{1}{p^2}.
\]
The vertices can be easily deduced from the non-quadratic terms in the
action (\ref{actionn}).  Each vertex carries a factor of $1/g^2$ and
each propagator carries a factor of $g^2$.

We use the position-space propagators in $2\omega$-dimensions.
These can be deduced from the Fourier transform
\begin{equation}
\int \frac{d^{2\omega}p}{(2\pi)^{2\omega} } \frac{ e^{ip\cdot x} }{ [p^2]^s }
=\frac{\Gamma(\omega-s)}{4^s\pi^\omega\Gamma(s) }\frac{1}{[x^2]^{\omega-s}}.
\label{ft}
\end{equation}
By setting $s=1$ we find the Green function in $2\omega$ dimensions:
\begin{equation}
\Delta(x)= \frac{ \Gamma(\omega-1)}{4\pi^\omega}\frac{1}{ [x^2]^{\omega-1} }
\qquad\text{which satisfies}\qquad
-\partial^2\Delta(x)=\delta^{2\omega}(x).
\end{equation}

We record some formulae which are useful in reproducing the computations 
in this paper.  Euler's $B$ function is defined as
\[
B(\mu,\nu)=
\int_0^1dx\,x^{\mu-1}(1-x)^{\nu-1}=\frac{ \Gamma(\mu)\Gamma(\nu)}{\Gamma(\mu+\nu)}
\]
and the $\Gamma$ function is
\[
\Gamma(n+1)=\int_0^\infty dt\,t^{n} e^{-t},
\]
which satisfies $\Gamma(n+1)=n!$, $\Gamma(1/2)=\sqrt{\pi}$, and the
combinatorial formulae
\begin{gather*}
\Gamma\bigl((2n+1)/2\bigr)=(n-1/2)(n-3/2)\dots(1/2)\sqrt{\pi},\\
\Gamma(n)\Gamma(1/2)=2^{n-1}\Gamma(n/2)\Gamma\bigl((n+1)/2\bigr).
\end{gather*}
One loop integrals in $2\omega$ dimensions can be computed using the dimensional regularization formulae~\cite{Ram89}:
\begin{gather*}
\int d^{2\omega}k (k^2+2p\cdot k+m^2)^{-s}=\pi^{\omega}\frac{
\Gamma(s-\omega)}{\gamma(s)}\left( m^2-p^2\right)^{\omega-s}\\
\int d^{2\omega}k k_\mu(k^2+2p\cdot k+m^2)^{-s}=
-p_\mu\pi^{\omega}\frac{ \Gamma(s-\omega)}{\gamma(s)}\left( m^2-p^2\right)^{\omega-s}\\
\begin{split}
\int d^{2\omega}k\,k_\mu k_\nu(k^2+2p\cdot k+m^2)^{-s}&=
\pi^{\omega}\frac{1}{\Gamma(s)}(m^2-p^2)^{\omega-s}
\\&
\qquad\times
\bigl[
p_\mu p_\nu\Gamma(s-\omega)-\tfrac{1}{2}g_{\mu\nu}\Gamma(s-\omega-1)(p^2+m^2)\bigr]
\end{split}
\end{gather*}
and the Feynman parameter formula,
\[
\prod_i A_i^{-n_i}=\frac{ \Gamma(\sum n_i)}{\prod_i\Gamma(n_i)}
\int_0^1 dx_1\cdots dx_k\,x_1^{n_1-1}\cdots x_k^{n_k-1}
\frac{ \delta(1-\sum x_i) }{\bigl[\sum_i A_i x_i\bigr]^{\sum n_i}}.
\]

\end{fmffile}

\end{document}